\documentclass[aps,prl,twocolumn,superscriptaddress]{revtex4}
\usepackage{subfigure}
\usepackage{graphicx}
\usepackage{rotating} \usepackage{color}
 \usepackage{amsmath,amsthm}
\usepackage{epstopdf}
\begin{document}
\title{Unexpected swelling of stiff DNA in a polydisperse crowded environment}  
\author{Hongsuk Kang}
\affiliation{Chemical Physics and Biophysics Programs, Institute of Physical Science and Technology, University of Maryland, College Park 20742, USA}
\author{Ngo Minh Toan}
\affiliation{Chemical Physics and Biophysics Programs, Institute of Physical Science and Technology, University of Maryland, College Park 20742, USA}
\author{Changbong Hyeon}
\thanks{hyeoncb@kias.re.kr}
\affiliation{Korea Institute for Advanced Study, Seoul 130-722, Korea}
\author{D. Thirumalai}
\affiliation{Chemical Physics and Biophysics Programs, Institute of Physical Science and Technology, University of Maryland, College Park 20742, USA}
\affiliation{Korea Institute for Advanced Study, Seoul 130-722, Korea}
\affiliation{Chemistry and Biochemsitry, University of Maryland, College Park 20742, USA}

\begin{abstract}
 We investigate the conformations of DNA-like stiff chains, characterized by contour length ($L$) and persistence length ($l_p$), in a variety of crowded environments containing monodisperse
soft spherical (SS) and spherocylindrical (SC) particles, mixture of SS and SC, and a milieu mimicking the composition of proteins in \emph{E. Coli.} cytoplasm.  
The stiff chain, whose size modestly increases in SS crowders up to $\phi\approx 0.1$, is considerably more compact      
at low volume fractions ($\phi \leq 0.2$) in monodisperse SC particles than in a medium containing SS particles.  
A 1:1 mixture of SS and SC crowders induces greater chain compaction than the pure SS or SC crowders at the same $\phi$ with the effect being highly non-additive.  
We also discover a counter-intuitive result that polydisperse crowding environment, mimicking the composition of a cell lysate, swells the DNA-like polymer, which is in stark contrast to the size reduction of flexible polymer in the same milieu. Trapping of the stiff chain in a fluctuating tube-like environment created by large-sized crowders explains the dramatic increase in size and persistence length of the stiff chain.
In the polydisperse medium, mimicking the cellular environment, the size of the DNA (or related RNA) is determined by $L/l_p$. 
At low $L/l_p$ the size of the polymer is unaffected whereas there is a dramatic swelling at  intermediate value of $L/l_p$. 
We use these results to provide insights into recent experiments on crowding effects on RNA, and also make testable predictions. 
\end{abstract}


\maketitle


The recognition that the crowded cellular environment can profoundly influence all biological processes, such as gene expression \cite{Morelli11BJ,Tabaka14NAR,Ge11PlosOne}, protein 
\cite{Zhou08ARB,Elcock10COSB,Cheung13COSB} and RNA folding  \cite{Denesyuk11JACS,Kilburn10JACS,Pincus08JACS} and protein-protein interactions \cite{Schreiber09ChemRev}, is getting increasing attention recently  although its importance was recognized decades ago \cite{Lerman71PNAS}. A simple calculation using the typical concentration of macromolecules shows that the average spacing between proteins in the {\it E. coli}  is $\sim 4$ nm, which is comparable to the size (radius of gyration) $R_{g} \approx 0.3N^{1/3}$nm of a folded protein \cite{Dima04JPCB} with $N\sim300$ amino acid residues. Therefore, the cellular interior, replete with macromolecules of different sizes and shapes (a polydisperse soup), is crowded, affecting the stability and shapes of the molecules of life. 
For example, compaction of DNA, relevant in a variety of biological processes ranging from organization of nucleoid in bacteria to DNA packaging in phage heads, is greatly facilitated in the presence of neutral osmotic agents. The effect of neutral polymer (poly-ethylene oxide) (PEO) in compacting DNA ($\Psi$-condensation) was demonstrated in a pioneering study by Lerman \cite{Lerman71PNAS}, who  showed that DNA undergoes a dramatic reduction in size if the concentration of PEO exceeds a critical value. Because the interactions between DNA and PEO in these experiments was established to be repulsive \cite{Hyeon06JCP_2}, $\Psi$-condensation is determined solely by the volume excluded to DNA by the crowding particles. Subsequently, Post and Zimm \cite{Post79Biopolymers} produced insightful theoretical explanations based on further experiments on crowding-induced compaction of DNA. 
In the intervening years a number of theoretical and experimental studies have explored various aspects of DNA compaction \cite{Grosberg82Biopolymers,Castelnovo04Macromol,ramos2005JPCB}.

Despite these advances a molecular description of how macromolecular crowding, especially a milieu containing polydisperse crowders, affects the spatial organization of DNA is poorly understood. 
Simulations of DNA in the presence of explicit crowding particles are computationally intensive, but their importance in describing the shapes of flexible polymers has been demonstrated in a number of recent studies \cite{kudlay2012JPCB,Kang15PRL,kim2015SoftMatter,shin2015softmatter,Shendruk15BJ}.  
Because of technical complications theories do not take into account the effects of polydispersity \cite{frisch1979JPS, naghizadeh1978PRL, Grosberg82Biopolymers,Thirumalai88PRA,Vasilevskaya95JCP,VanderSchoot1998Macromolecules,Krotova2010PRL,Diamont00PRE,Castelnovo04Macromol,Shaw91PRA}. 
In addition, the extent of structural changes and how they depend on mixtures of crowding particles of different shapes for realistic sizes of crowders found in the cytoplasm are unknown. 
Most of the studies have focused on intrinsically flexible polymers with very little work on relatively short stiff DNA-like polymers, and related RNA. 
Stiff short chains can display quantitatively different behavior than the one associated long flexible polymers, as argued in the case of cyclization of DNA \cite{HyeonJCP06,Vafabakhsh12Science,le2014NAR}.  
These issues take on added importance because of potential relevance to genome confinement \cite{lieberman09Science,parry2014Cell,Zhou08ARB}, where the structural organization is functionally related to gene expression.  

Inspired by these observations,  we first performed simulations of worm-like chain \cite{Marko95Macro} 
(WLC - a reliable polymer model for describing many of the properties of DNA at high salt concentration) in monodisperse solutions containing soft spherical (SS) and spherocylindrical (SC) crowding particles. 
Here, we focus on DNA-like chains in the limit where the contour length ($L=N_ml_0$, $N_m$ is the number of monomers, and $l_0$ is the bond length) is not significantly longer than the persistence length, $l_p$. 
The systematic study leads to a number of unexpected predictions, which can be tested using synthetic polymers in the presence of nanoparticles and DNA using crowding agents. 
(i) The SC crowders induce greater compaction than SS particles. 
The compaction is accompanied by substantial reduction of $l_p$ (by nearly a factor of two) similar in magnitude to that observed by ion-induced compaction of DNA and RNA \cite{moghaddam09JMB}. 
(ii) One of the most striking results of our study is that stiff chains are more compact in a mixture of SS and SC, at physiologically relevant volume fractions, than in monodisperse crowders at the same volume fraction. The substantial compaction of DNA in the mixture is due to ordering (dense packing) of the SS crowders around DNA due to depletion attraction induced by the SCs.
(iii) We also carried out simulations of DNA in polydisperse crowders mimicking the composition of macromolecules in 
\emph{E. coli}.  
Surprisingly, we obtained a counter-intuitive and theoretically unanticipated result that polydisperse spherical crowding particles cause swelling of DNA at volume fraction $\approx$ 0.3 appropriate for \emph{E. Coli.} cytoplasm. 
Surrounded predominantly by large-sized crowders, the swollen conformation of DNA (nearly three times increase in volume of the chain) is entropically driven.

\section{Results}

\begin{figure}[t]
\begin{tabular}{c}
\includegraphics[width=1.0\columnwidth]{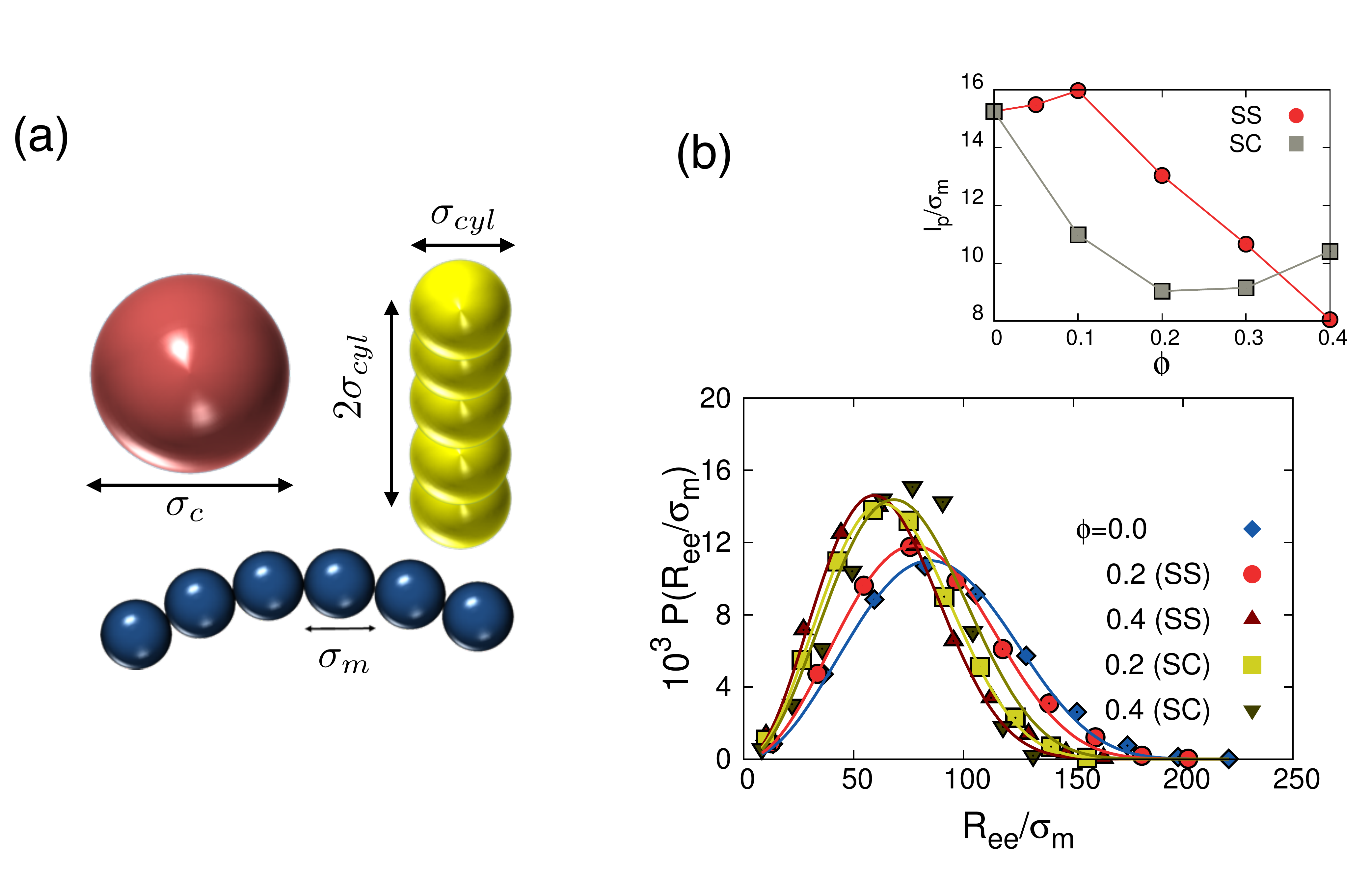}
\end{tabular}
\caption{\label{Fig1}
(a) Coarse-grained models of spherical (SS) crowder (red), spherocylinderical (SC) crowder (yellow) and WLC chain (blue) used to model DNA. The relevant dimensions are labeled $\sigma_{sph}$, $\sigma_{cyl}$, and $\sigma_m$. 
(b) End-to-end distance distribution of WLC at two different volume fractions ($\phi=0.2$, 0.4) of SS and SC crowders. 
The top panel shows the persistence length ($l_p$) of the polymer as a function of the volume fraction of SS and SC crowders. 
}
\end{figure}

\begin{figure*}[t]
\begin{tabular}{c}
\includegraphics[width=1.8\columnwidth]{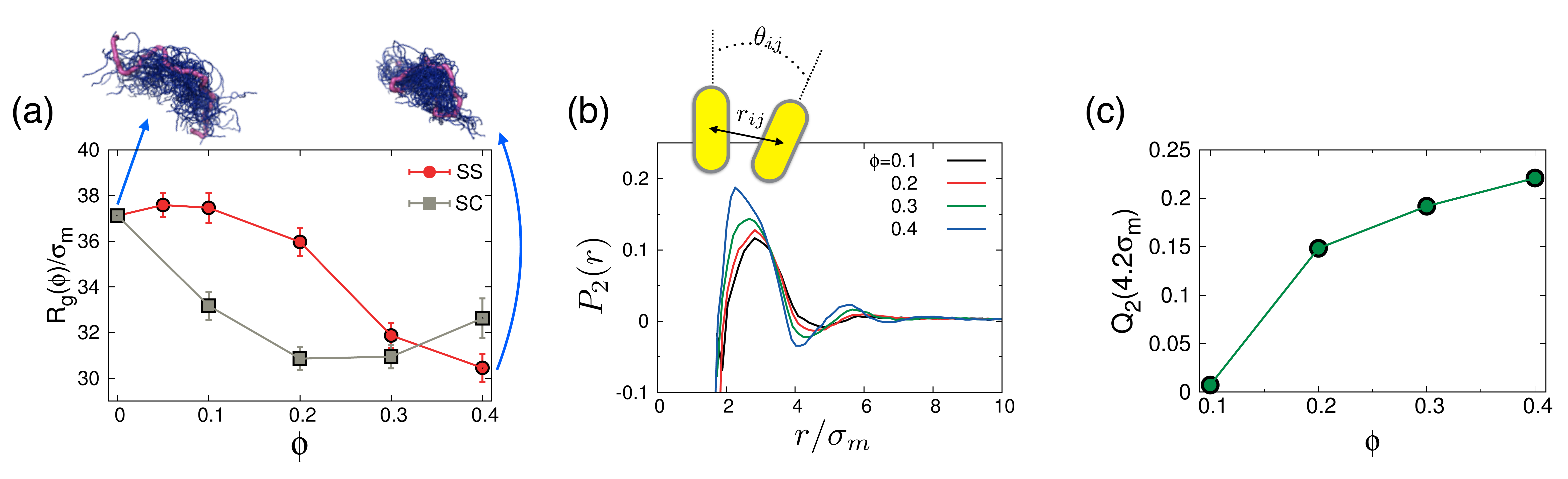}
\end{tabular}
\caption{\label{Fig1-2}
(a)  The change of $R_g$ of the polymer from $R_g(0)=37.6$ nm with increasing $\phi$. 
(b) Liquid crystal order parameter for SC crowders, averaged over the ensemble of crowder particles, as a function of distance.  
(c) The extent of local nematic ordering is quantified with $Q_2(r)$ at $r=4.2\sigma_m$. 
}
\end{figure*}

{\bf Crowding-induced softening of DNA. }
In order to set the scale for bending energy of the polymer, we first calculated the end-to-end distance distribution, $P(R_{ee})$, in the absence of crowders to obtain the bare persistence length, $l_p$ for $\phi=0$. 
By fitting the simulated $P\left(r \right)$ to an analytic expression for WLC (Eq.S3) \cite{HyeonJCP06}, we obtained $l_{p}=\left(15.4 \pm 0.08\right) \sigma_{m}\approx49$ $\mathrm{nm}$ using $\sigma_{m}\approx3.18$ nm (\ref{Fig1}b), which coincides with $l_p$ for DNA in high monovalent salt concentration.  
Because $l_{p}/\sigma_{m}>1$, the shape of the chain in the presence of crowding particles should be determined by an interplay of bending rigidity and attractive depletion interaction due to crowders. 
$P(R_{ee})$ of the WLC with varying $\phi$ for both monodisperse SS and SC crowders show gradual shift to the smaller $R_{ee}$ with increasing $\phi$ from 0 to 0.4 (\ref{Fig1}b).   
At high $\phi$, as $\phi$ increases, $l_p$ obtained from the fit of $P(r)$ to Eq.S3 gradually decreases for both SS and SC crowders, 
implying that the crowding particles induces compaction of the WLC polymer (\ref{Fig1}b, top panel).
Interestingly, the WLC polymer exhibits a non-monotonic dependence of size with increasing $\phi$. 
The SS crowders induce a minor increase of $l_p$ (stiffening or expansion) of the chain (\ref{Fig1}b top panel and \ref{Fig1-2}a) for $0<\phi\leq 0.1$, followed by a decrease of $l_p$ (softening or compaction) at larger $\phi=0.2-0.4$.  
In contrast, the SC crowders reduces $l_p$ much more efficiently than the SS crowders 
at $\phi\leq 0.3$ in that an expansion similar to the one in the SS crowders at $\phi\approx 0-0.1$ is not observed. 
Instead, there is a modest re-stiffening of the DNA due to SC crowders when $\phi$ is in the range from $\phi=0.2$ to $\phi=0.4$. 
We further substantiate this result below by calculating the change in the polymer size ($R_g$) and nematic order parameter of the crowders. 
The slight increase in $l_p$ not withstanding, the overall trend is that there is substantial softening ($l_p$ decreases by nearly a factor of 2) as $\phi$ increases from 0 to 0.4.\\

{\bf Dependence of $R_{g}$ on $\phi$ for monodisperse crowders.}
Snapshots of polymer conformations at different values of $\phi$ show modest compaction as $\phi$ increases (\ref{Fig1-2}a). 
Similar to $l_p$, the dependence of $R_{g}$ on $\phi$ for SS and SC crowders displays substantial difference. 
At $\phi=0.2$, $R_{g}(0.2)$ is smaller than $R_g(0)$ by only 4 \% in SS crowders whereas $R_{g}(0.2)$ decreases by 17 \% in SC crowders. 
Given that the volume of the chain is $\sim R_{g}^{3}(\phi)$ the extent of compaction induced by SC is substantial compared with SS crowders. 
The quantitative difference between the effects of SS and SC crowders on the chain compaction is explained using the depletion interaction (or Asakura-Oosawa (AO) interaction \cite{Asakura58JPS}) that produces an effective attraction between monomers. 
The strength of the AO interaction for SS, roughly given by $\approx\phi k_BT/\sigma_{sph}^2$, has to exceed the energy ($\sim \frac{1}{2} (L/l_p) k_{B} T$) required to bend the polymer on scale $l_{p}$ for compaction to occur.
For small $\phi$, it is unlikely that the AO attraction can compensate for the bending penalty. Thus, we expect little change in $R_{g} \left( \phi \right)$ at small $\phi$ for SS crowders. 
On the other hand, the strength of the AO interaction on the WLC for SC crowders is $\approx \phi P (\sigma_{m}/\sigma_{cyl}^2)k_BT$
where $P(=2\sigma_{cyl})$ is the cylinder length. 
In both cases, the origin of the AO depletion interaction, which has to exceed the bending energy to compact the stiff chain, leading to an effective short range (on length scale $\sim \sigma_{m}$) attraction between monomers and polymer compaction, is purely entropic. 
\ref{Fig1-2}a shows that $R_g(\phi)$ decreases monotonically till $\phi\approx 0.3$ for SC, indicating that for the parameter used, the strength of the attractive AO interaction due to SC crowders can exceed the penalty for bending the chain on scale $\sim l_p$.

The strong effect of compaction of the WLC chain induced by SC relative to SS crowders can also be quantified by comparing the volume excluded to the polymer by the crowders. On the basis of scaled particle theory, we can estimate the entropy cost of inserting a hard sphere of dimension $\sigma_{HS}$ in a box containing hard fluid particles. The entropy difference for inserting the hard sphere of diameter $\sigma_{HS}$ is related to 
$C(\sigma)$ \cite{Ogston70JPC},
\begin{equation}
C(\sigma_{HS})=\frac{V_{\mathrm{cyl}}}{V_{\mathrm{sph}}}=\frac{3\left[(\sigma_{HS}+\sigma_{\mathrm{cyl}})^{2} P +(\sigma_{HS}+\sigma_{\mathrm{cyl}})^{3}\right]}{2\left(\sigma_{HS}+\sigma_{\mathrm{sph}}\right)^{3}}
\end{equation}
where $V_{cyl}$ ($V_{sph}$) is the volume excluded by rod-like (spherical) crowders. For the parameters listed in Table S1, we find that $V_{cyl}>V_{sph}$ provided $\sigma_{HS} \approx R_{g} > \sigma_{cyl}$ or $\sigma_{sph}$. The entropic cost of inserting a spherical particle of size $R_{g}$ into a fluid of cylindrical crowders exceeds that for inserting it into a system consisting of spherical crowders. By achieving greater compaction of the WLC in SC crowders, the entropy difference is minimized, thus, explaining the results in \ref{Fig1-2}a.

The pattern of compaction in the two crowding environment is qualitatively different. 
For SS crowders, compaction occurs only when $\phi$ exceeds $\approx0.1$ ($dR_{g}/d\phi \approx 0$ as $\phi \rightarrow 0$).
Although the trend is not so clear as in $l_p(\phi)$ (\ref{Fig1}b, top panel), $R_g(\phi)$ (\ref{Fig1-2}a) shows a signature of minor swelling in the range of $\phi=0-0.1$, which was also observed in experiments on the effects of crowding on a ribozyme (see below) \cite{Kilburn10JACS}.     
In contrast, for anisotropic crowding agents (SC crowders), 
$R_{g}\left( \phi \right)$ decreases monotonically for $\phi \leq 0.3$ ($dR_{g}/d\phi<0$ for all $\phi$) and increases from $\phi=0.3$ to $\phi=0.4$.
High volume fraction of SC results in the reswelling of WLC (Figure 2a), showing that the shape of the crowding particles can have a profound effect on size of a stiff chain. 
\\

{\bf Local nematic order and increase in $R_{g}$. }
Interestingly, as $\phi$ exceeds 0.3, $R_{g} \left( \phi \right)$ of the chain increases in SC crowders (\ref{Fig1-2}a). 
The increase in $R_g(\phi)$ at higher $\phi$ is due to plausible development of local nematic ordering for SC crowders at $\phi>0.3$. 
The system consisting of pure SC crowders will undergo an isotropic to nematic phase transition if $\phi$ exceeds a critical value $\phi_\mathrm{I \rightarrow N}$. We calculated the liquid crystal order parameter, $\langle P_{2}(\cos \theta) \rangle$, where $\theta$ is the angle between the long axes of any pairs of SC crowders, $P_{2}(x) = \frac{1}{2} \left(3 x^{2} - 1 \right)$ is the second Legendre polynomial, and $\langle\ldots\rangle$ denotes the ensemble average. $\langle P_{2}(\cos{\theta}) \rangle$ is almost zero $\left(<0.05\right)$ for all $\phi$, which means that even the highest $\phi$($=0.4$) is less than $\phi_\mathrm{I \rightarrow N}$. 
However, locally the crowders adopt nematic-like state. 
To ascertain if this is the case, we calculated $P_{2}(r) =\langle \sum_{i,j} P_{2} \left( \cos \theta_{ij} \right) \delta \left( \left| \vec{r}_{i} - \vec{r}_{j} \right| - r \right)\rangle$ as a function of distance, $r$.
\ref{Fig1-2}b shows that the angular correlation of SCs becomes stronger as $\phi$ increases at short distances.

To quantify the extent of ``local'' nematic ordering, we calculated $Q_2(r)=N_{P}^{-1} \int_{r_{\text{min}}}^{r} P_{2} \left( r' \right) \mathrm{d} \vec{r'} $, where $N_{P}$ is the number of pairs separated by $r$ and $r_{\text{min}}$ is the minimal distance between SC crowders. 
At the distance $r=2.3\sigma_{m}$ where the pair correlation has first peak for $\phi=0.4$ (\ref{Fig1-2}b), $P_{2}(2.3\sigma_m)$ increases significantly  from 0.0 to 0.23 as $\phi$ increases from 0.1 to 0.4  (\ref{Fig1-2}c). 
At $r=4.2 \sigma_{m}$, $Q_{2}(r)$ $\approx 0.23$ suggests that the local nematic ordering of cylindrical crowders is reinforced in the vicinity of the WLC. 
This result implies that relatively stiff chain induces ordering of SC crowders along the polymer axis, and strengthens the anisotropic interaction of rod-like particles. At high $\phi$ the chain may be thought of as being in a local nematic field, which elongates the polymer along the local direction of the nematic field, thus explaining the increase in $R_g(\phi)$ when $\phi$ exceeds 0.3. 
\\

\begin{figure}[ht]
\includegraphics[width=1.0\columnwidth]{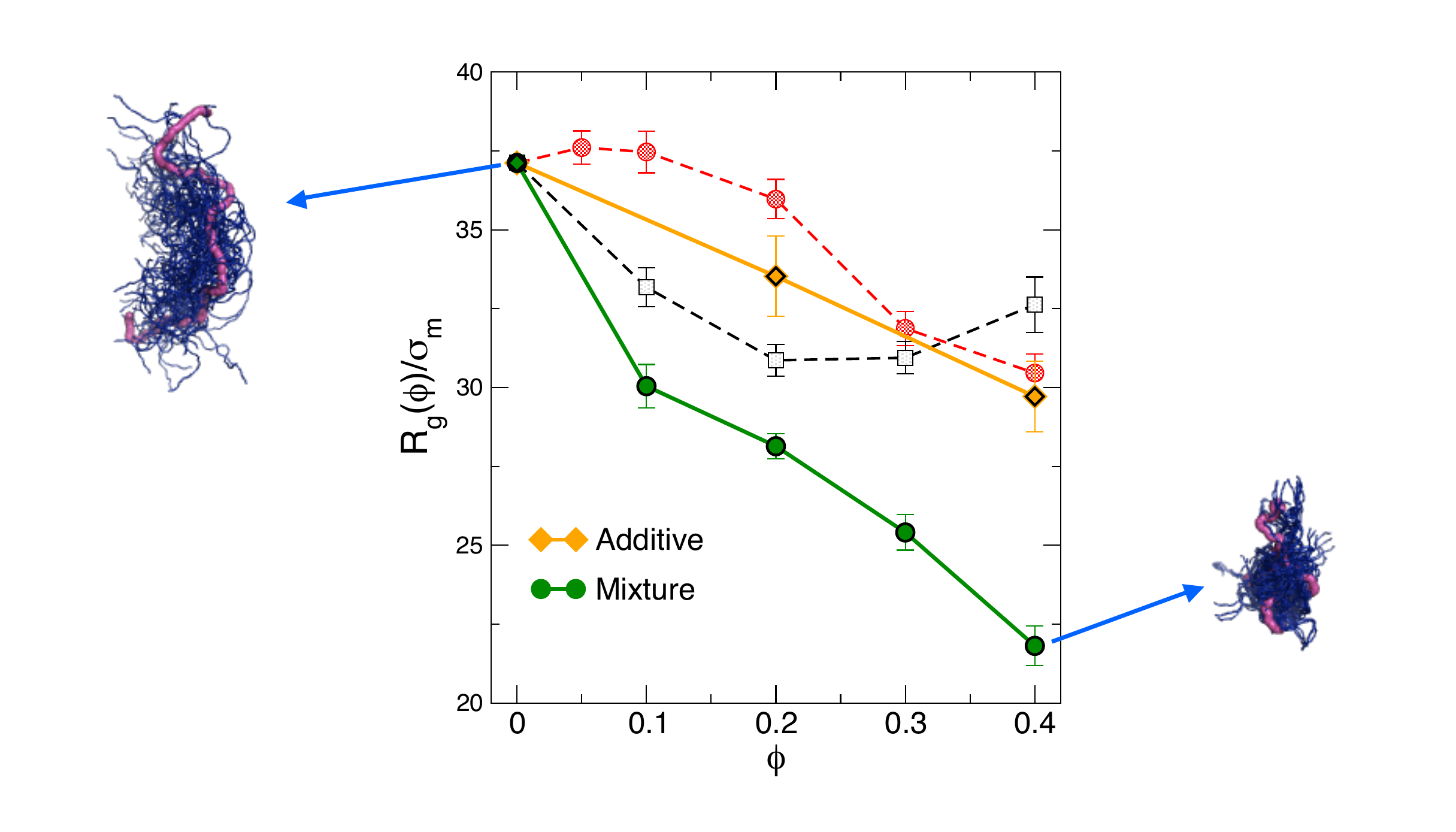}
\caption{\label{mixture-1}
Compaction of stiff chain in a mixture of SS and SC crowders. 
$R_g$ of WLC as a function of $\phi$ of 1:1 SS and SC mixture (green). The orange line is the calculated $R_g$ at each $\phi$ value by assuming that the effects of SS and SC crowders on WLC compaction are additive. 
The results of monodisperse SS and SC crowders in \ref{Fig1-2}a are shown with the dashed lines to underscore the substantial enhancement of the chain compaction by the mixture.     
}
\end{figure}

\begin{figure}[ht]
\includegraphics[width=1.0\columnwidth]{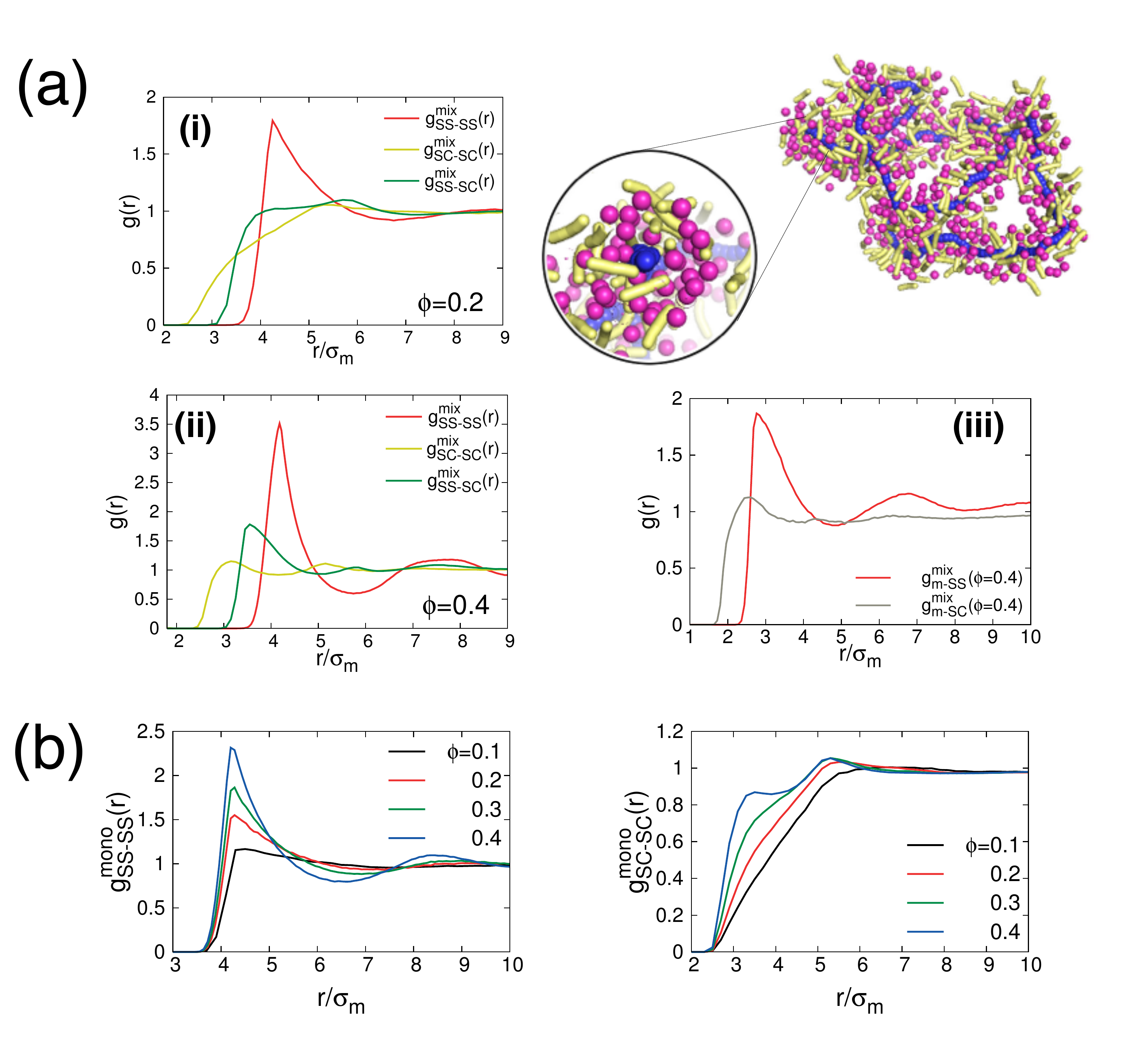}
\caption{\label{mixture-2}
(a) RDFs are calculated for all possible combinations of crowder-crowder pairs (at (i) $\phi=0.2$ and (ii) 0.4) and (iii) monomer-crowder (at $\phi=0.4$). 
Distribution of SS and SC crowders around the WLC chain is depicted using a snapshot from the simulations (The particle sizes are not in scale. Although $\sigma_{sph}=4\sigma_m$ and $\sigma_{cyl}=4^{2/3}\sigma_m$ were used in actual simulations, we deliberately reduced the sizes of SS and SC crowders for clear illustration of the crowders around the chain).
(b) RDFs of SS-SS and \text{SC-SC} pairs at $\phi=0.1-0.4$ in monodisperse condition are presented to highlight the enhanced local ordering of SS and SC crowders in the mixture.  
}
\end{figure}

{\bf Non-additive effect in a mixture of spheres and spherocylinders. }
The dependence of $R_{g} \left( \phi \right)$ on $\phi$ of the WLC in the 1:1 mixture of SS and SC crowders is shown in \ref{mixture-1}. 
The mixture has a profound effect on the size of DNA. 
The value of $R_g(0.4)$ is reduced by over 40\% from $R_{g}(0)$, 
whereas the maximum compaction in monodisperse SC at $\phi=0.4$ is only 10\%.
In order to highlight the non-additive effects of the two crowders with different shape, we also show, at $\phi=0.2$ and $\phi=0.4$, the expected result for $R^{A}_g(\phi)=R_g(0)+\delta R^{A}_g(\phi)$, where $\delta R^{A}_g(\phi)= \delta R_{g}^{\text{SS}}(\phi/2) + \delta R_{g}^{\text{SC}}(\phi/2)$, obtained using the assumption that the effect of crowders on the extent of $R_g$ reduction is additive. 
Remarkably, $R_{g}(\phi)$ is significantlly lower than $R_{g}^{A}(\phi)$ (\ref{mixture-1}), (or 
$|\delta R_g(\phi)|>|\delta R_g^A(\phi)|$, $|\delta R_g^{\text{SC}}(\phi)|$, $|\delta R_g^{\text{SS}}(\phi)|$), 
indicating that the mixture of SS and SC restricts the volume available to the WLC to a much greater extent than the individual components do.

The surprising finding of significant compaction in the mixture can be qualitatively explained using the notion of depletion potentials for mixture of SS and SC. Consider the interaction between two spherical particles in the presence of rods (SC particles). If a SC particle is spatially trapped then the rod loses translational and rotational entropy because of orientational restrictions.
The large unfavorable entropy loss results in the depletion of the SC from the space where the SC particle is trapped. The result is that there would be an excess osmotic pressure due to the AO attraction that pushes the SS and the monomers together. Two consequences of the entropy-driven depletion interactions are : 
(i) Due to the attractive interactions, the SS particles are more closely packed than in the absence of the SC 
(compare $g^{\text{mix}}_{\text{SS-SS}}(r)$ at $\phi$ from \ref{mixture-2}a and $g^{\text{mono}}_{\text{SS-SS}}(r)$ at $\phi/2$ from \ref{mixture-2}b (monodisperse crowders)). Indeed, in the limit of $\sigma_{cyl}<P<\sigma_{sph}$, it has been experimentally shown that the addition of a small fraction (by volume) of the SC can even lead to crystallization of low density suspension of hard spheres \cite{Vliegenthart1999JCP,oversteegen2004JPCB}. 
(ii) We also expect that the excess volume available to the WLC should be greatly reduced compared to the monodisperse crowders. In such a confined space the WLC should be considerably more compact than in the presence of monodisperse crowders at the same $\phi$.

The expected enhancement in the packing of the SS due to the SC is evident in the pair distribution function $g(r)$. 
The results for $\phi=0.2$ and $\phi=0.4$ (\ref{mixture-2}a-(i), (ii)) show that both the radial distribution functions (RDFs), $g^{\text{mix}}_{\text{SC-SC}} \left( r \right)$, between SCs and $g^{\text{mix}}_{\text{SS-SC}} \left( r \right)$, between SC and SS, do not exhibit significant structure. 
In sharp contrast, $g^{\text{mix}}_{\text{SS-SS}} \left( r \right)$ has the structure corresponding to a high density liquid (especially at $\phi=0.4$ in \ref{mixture-2}a-(ii)), which is remarkable given that at $\phi/2(=\phi_{SS}=0.2)$, corresponding to the same volume fraction occupied by SS, $g^{\text{mono}}_{\text{SS-SS}}(r)$ (\ref{mixture-2}b) is relatively featureless.

The much stronger depletion force due to the SC crowders results in a considerable reduction in the volume accessible to the chain, which explains the dramatic reduction of $R_{g} \left( \phi \right)$ compared to the pure component case.
The WLC at $\phi=0.4$ is surrounded predominantly by SS crowders, which is shown by $g^{\text{mix}}_{\text{m-X}}(r)$, the RDF between the monomer and the crowders ($X=$ SS or SC) (\ref{mixture-2}a-(iii)). 
The number of SS near the monomer calculated using $N_{SS}=4 \pi  \left( \frac{N_{0}}{V} \right) \int_{0}^{r_{min}} r^2 g_{\text{m-SC}} \left( r \right) dr$ where $r_{min}$ is the first minimum in the $g_{\text{m-SS}}\left( r \right)$ at $\phi=0.4$ is $11.4$. A similar calculation for SC gives $N_{\text{SC}}=4.7$.
\\

{\bf Polydisperse crowders mimicking the \emph{E. coli} cytoplasm composition. }
To a first approximation the \emph{E. coli}  cytoplasm may be represented by a mixture of spheres because majority of the crowders present in large numbers (ribosomes, polymerases and other large complexes, and smaller particles) are compact \cite{Roberts2011PLOSCOMPBIO}. In order to assess the shape of DNA-like chain in such a mixture, we investigated the effect of polydisperse spherical particles on the conformational fluctuations of the stiff polymer.

\begin{figure}[b]
\centering{
\includegraphics[width=1.0\columnwidth]{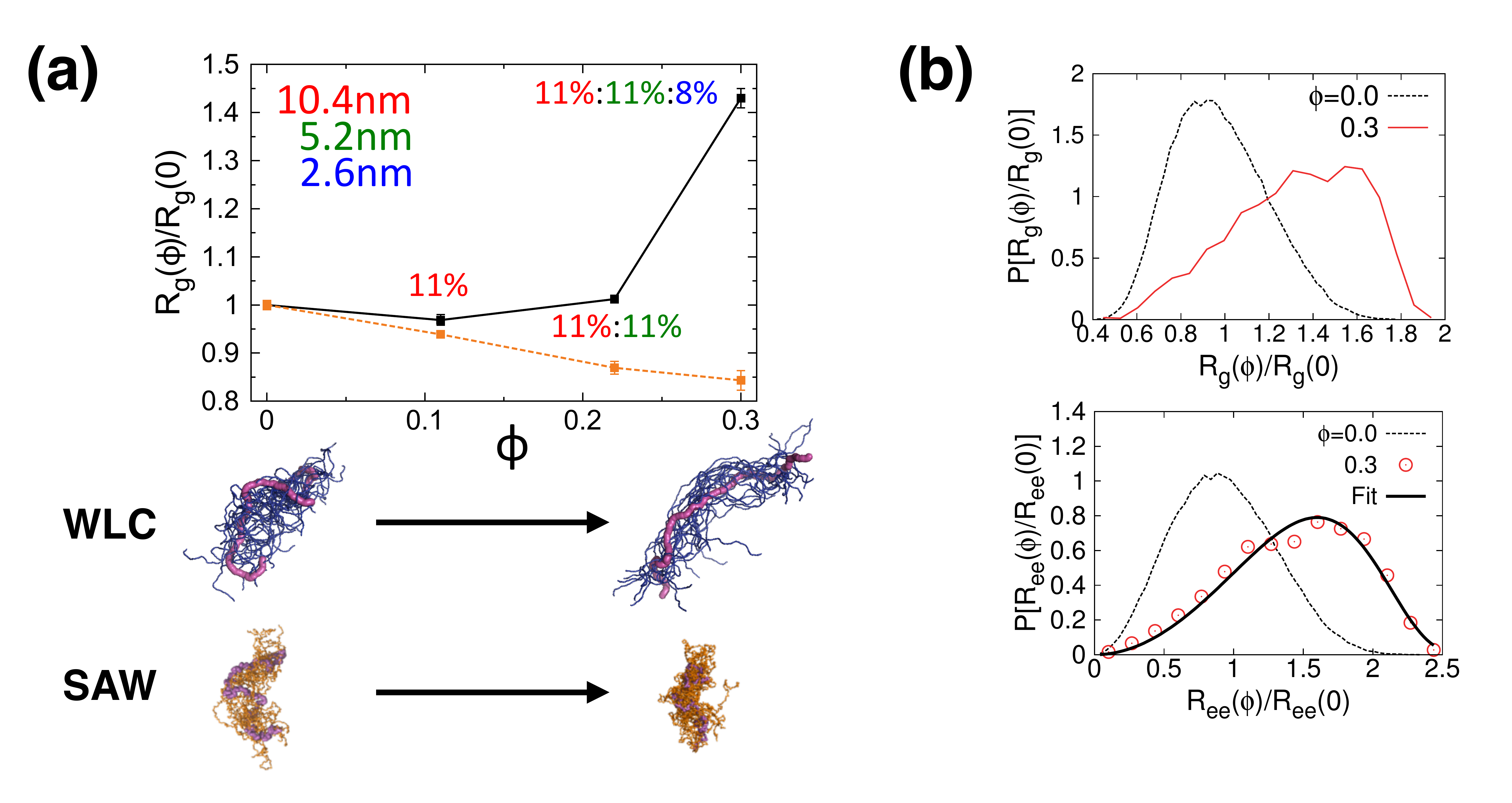}
}
\caption[Crowding Effects of polydisperse SS on DNA]
{\label{fig:EcoliCrowding1} 
Effects of polydisperse soft sphere mixture on the size of WLC ($L/l_p=20$ at $\phi=0$). 
(a) $R_{g}$ is calculated for WLC (black solid line) and SAW (orange dashed line) polymers at (i) $\phi=$ 0.11 of crowders with $r_1=10.4$ nm, (ii) $\phi=0.22$ with $r_1=10.4$ nm (11 \%) and $r_2=5.2$ nm (11 \%), and (iii) $\phi=0.3$ with $r_1=10.4$ nm (11 \%),  $r_2=5.2$ nm (11 \%),  $r_3=2.6$ nm (8 \%).  
The largest crowder represents the ribosomes, the 5.2 nm crowders correspond to polymerases and other large protein complexes, and the smallest sized particles is the average size of other crowders in the milieu.
Structural ensembles of polymers at $\phi=0$ and 0.3 are shown at the bottom, demonstrating the contrasting effect of polydisperse crowders on the conformations of WLC and SAW. 
(b) $P(R_g)$ (top) and $P(R_{ee})$ (bottom) of WLC at $\phi=0$ and $0.3$. 
The fit of $P(R_{ee})$ at $\phi=0.3$ to Eq. S3 in the text gives $l^{\phi=0.3}_p=41\sigma_m\approx 130$ nm, which is $\sim 2.6$ fold greater than $l_p^{\phi=0}(\approx 15\sigma_m\approx 49 nm)$. 
$P(R_g)$ of SAW (flexible self-avoiding chain) is shown in Figure S3.
}
\end{figure}

Strikingly, the behavior of the stiff chain in a polydisperse mixture of SS particles differs drastically from those in the mixture of SS and SC of the same volume (\ref{mixture-1}).
A few features in the non-monotonic dependence of $R_{g} \left( \phi \right)$ as a function of $\phi$ (\ref{fig:EcoliCrowding1}a) are worth pointing out. (i) There is a very modest reduction in $R_{g}\left( \phi \right)$ at $\phi \approx 0.1$, which has only large crowding particles. Such a decrease is comparable to that found in \ref{Fig1-2}a. 
(ii) Unexpectedly, $R_{g} \left( \phi \right)$ starts increasing in a mixture containing large and medium sized crowders. Most strikingly, in the mixture roughly mimicking the \emph{E. coli} cytoplasm \cite{Roberts2011PLOSCOMPBIO} there is a large increase ($\sim 40$\%) in $R_{g} \left( \phi \right)$ compared to $\phi=0$. The competition between bending free energy and depletion potential leading to a dramatic swelling of the stiff chain is counter-intuitive. The ensemble of the chain conformations (\ref{fig:EcoliCrowding1}a)
exhibiting the expansion of the chain  captures these effects visually. 
(iii) The \emph{E. coli} mixture dramatically stiffens the polymer. 
The persistence length of the chain for the swollen chain in the \emph{E. coli} mixture is about 2.6 times larger than for the one at $\phi=0$. 
The stiffening effect of mixture of spherical crowders on WLC captured the snapshots in \ref{fig:EcoliCrowding1}. (iv) It is noteworthy that in the polydisperse mixture with \emph{E. coli.} composition reduces the size of a flexible chain that lacks bending penalty (\ref{fig:EcoliCrowding1}a and Figure S3), underscoring the importance of chain stiffness.

\begin{figure}[ht]
\centering{
\includegraphics[width=1.0\columnwidth]{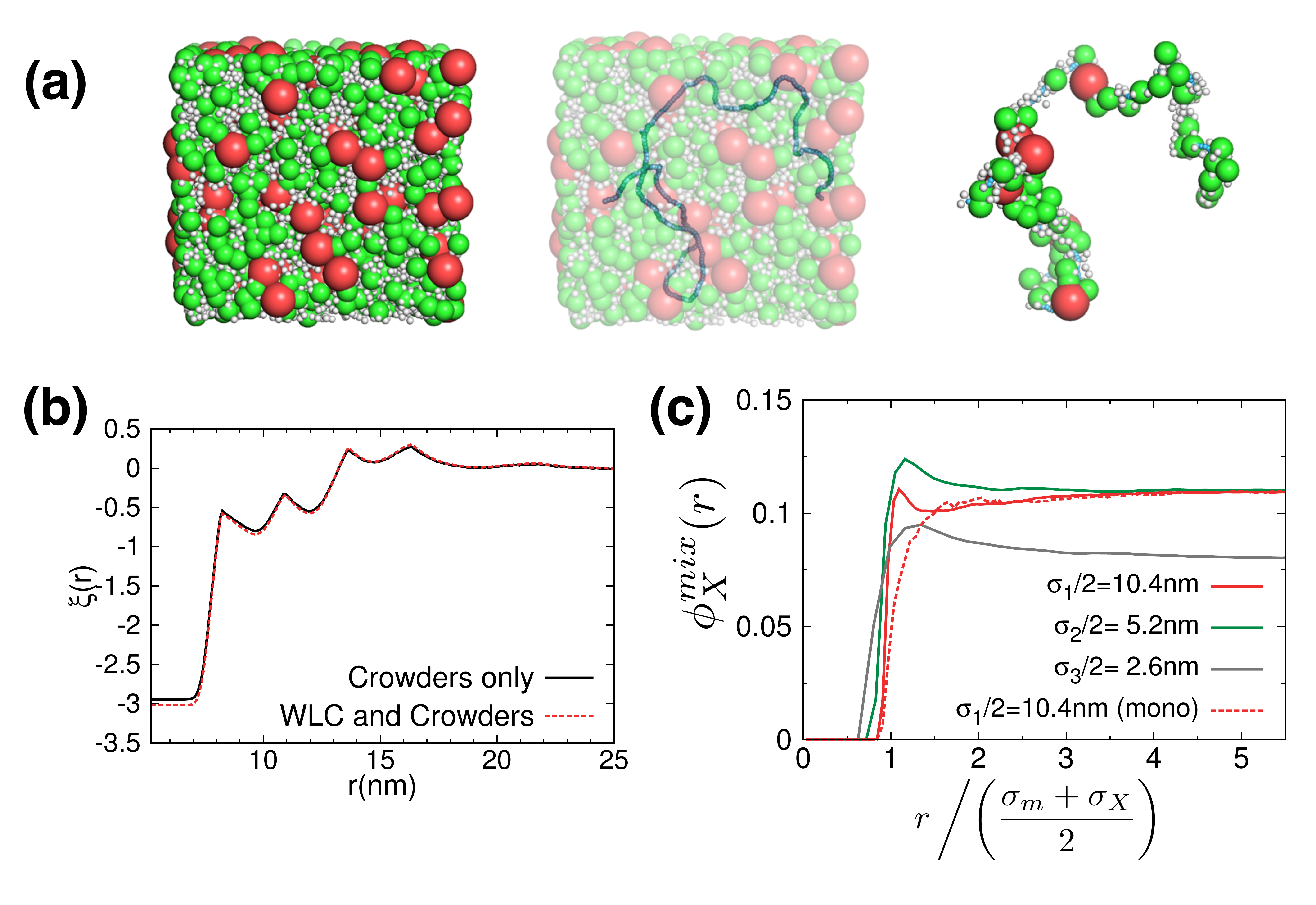}
}
\caption[Crowding Effects of polydisperse SS on DNA]
{\label{fig:EcoliCrowding2} 
(a) A snapshot from simulation, demonstrating 
(left) the polydisperse crowding environment, 
(middle) the WLC inside the crowders, and 
(right) the crowding particles decorating the monomers. 
(b) Local size ordering correlation function, $\xi(r)$, indicates a non-uniform,  heterogeneous size ordering of polydisperse crowders.
(c) $\phi^{\text{mix}}_X(r)$ ($X=1, 2, 3$), the fraction of volume around a monomer occupied by three different types of crowders in polydisperse crowding environment (case (iii) in (a)) as a function of monomer-crowder distance (solid lines). The dashed line shows the corresponding quantity for monodisperse crowding environment.   
}
\end{figure}

The reswelling at high $\phi$ (\ref{fig:EcoliCrowding1}a, \ref{fig:EcoliCrowding1}b) can be understood qualitatively using the following arguments. At a specified total volume fraction there are a lot more small crowding particles than large ones. Therefore, the entropy of the crowding particles is maximized if the WLC is surrounded by the larger sized particles with the smaller ones being further away from the chain. In this picture, the WLC is localized in a region in which the larger sized particles are with higher probability in proximity to the monomers (Figure S4).  
Because the interactions between the crowding particles and the stiff chain are repulsive the DNA chain would prefer to be localized in a largely crowder-free environment.  If we assume that such a region is roughly spherical, created predominantly by the largest crowders, then its size has to be on the order of $R_g^3$ to accommodate the WLC chain. In such a cavity there is an entropic cost to confine the stiff chain.  The probability of finding such a region decreases exponentially as $R_g$ gets large. In addition, in a spherical region the DNA would form spool-like structures requiring overcoming bending energy. The combination of these effects makes it likely that an optimal spherical regime can be found to minimize the free energy of WLC. If the region is cylindrical and large enough such that tight hairpins (costing substantial bending penalty) are avoided then the chain free energy may be minimized by confining it in a roughly cylindrical cavity. Such a possibility is supported by simulations, which show that
on an average the shape of the depletion zone is aspherical resembling a fluctuating tube (see  images in \ref{fig:EcoliCrowding2}a). 
As a result, we can visualize the polymer to be essentially confined to an anisotropic (but fluctuating) tube in which  transverse fluctuations of the chain are restricted but one in which tight hairpin turns cannot form because of bending penalty. 
In such a cavity  the chain stiffens, thus expanding in size in order to minimize both the bending penalty and entropy cost of confinement.
A more quantitative and accurate theory is difficult to construct because of the many body correlation among the \emph{polydisperse} crowding particles.

The plausibility of the physical picture given above is further substantiated by examining how the crowding particles with different sizes are arranged in space and how they surround the extended WLC.   
Distributions of the polydisperse crowding particles are not uniform, but exhibit local size ordering. 
This is evident in the local size correlation function ($r$-dependent size variance), 
$\xi(r)=\langle d_id_j\rangle_r-\langle d\rangle^2$ (\ref{fig:EcoliCrowding2}b) where 
$\langle d\rangle$ is the mean diameter and $\langle d_id_j \rangle_r$ in the first term denotes an average of the product of two diameters $d_i$ and $d_j$ taken over crowders ($j$) located at a distance $r$ from a crowder $i$  \cite{Williamson2013SoftMatter}. 
The local size correlation $\xi(r)$ (we set $\langle d_id_j\rangle_r=0$ when no pair exists), which is by definition zero for both monodisperse crowders and 
homogeneously distributed polydisperse crowders, 
reveals the presence of size ordering up to $r\leq 15-20$ nm.
This implies that there is a non-uniform ordering in the mixture of spherical particles, which is responsible for reswelling (\ref{fig:EcoliCrowding1}a). 

The fraction of volume occupied by three different crowders ($\phi^{\text{mix}}_X(r)\left(=\frac{4\pi}{3}\left(\frac{\sigma_X}{2}\right)^3\times \phi_Xg^{\text{mix}}_{m-X}(r)\right)$ with $X=1,2,3$) as a function of distance from the WLC monomers  in  \ref{fig:EcoliCrowding2}c further captures the non-uniform distribution of crowders. 
The crowders with intermediate size ($X=2$) occupy the largest volume near the WLC. 
In addition, the comparision of $\phi^{\text{mix}}_2(r)$ with $\phi^{\text{mono}}_2(r)$ (the volume fraction of monodisperse 11 \%-crowders around WLC monomer; the dashed line in \ref{fig:EcoliCrowding2}c) shows that 
the intermediate ($X=2$) and large ($X=1$) sized crowders are pushed closer to the monomers by the small crowders, which confines a segment of the polymer to a tube-like region (\ref{fig:EcoliCrowding2}a).
The depletion forces in a polydisperse solution give rise to a spatial inhomogeneity of crowders, resulting in the chain being confined to a cylindrical region created by the large-sized ($X=1,2$) crowders. 
The expansion of the chain in such a confined space \cite{deGennesbook} provides a plausible physical explanation for the large increase in the size of the DNA. 

\section{Discussion}
{\bf Swelling and collapse of DNA. }
The counter-intuitive finding that a stiff polymer, with $\gamma = L/l_p$ not large (see below), can swell relative in a polydisperse mixture of spheres, is (to our knowledge) unprecedented. There are simulation and theoretical studies predicting the collapse of flexible polymers and proteins in mixed solvents due to volume exclusion effects alone \cite{brochard1980Ferroelectrics,Xia2012JACS}. 
However, the present study shows precisely the opposite behavior for stiff chains whereas a flexible chain tends to become compact (not a globule in the {\it E. Coli.} like milieu). 
We propose that this unusual effect is related to an interplay of chain bending and the complex depletion effect in a polydisperse crowding system. In order to substantiate our proposal we carried out simulations for chains with $\gamma$ varying from 2 to 10 in the model {\it E. Coli.}-like system. The simulations show hardly any change in $R_g$ (see Figure S5). For all the values of $\gamma(\leq 10)$ the stiff polymer can be accommodated in a large enough region in which the crowding particles do not suppress the conformational fluctuations. Only when $\gamma$ exceeds a minimum value, but is not too large, then chain swelling occurs by formation of tight turns.

\begin{figure*}[t]
\includegraphics[width=1.5\columnwidth]{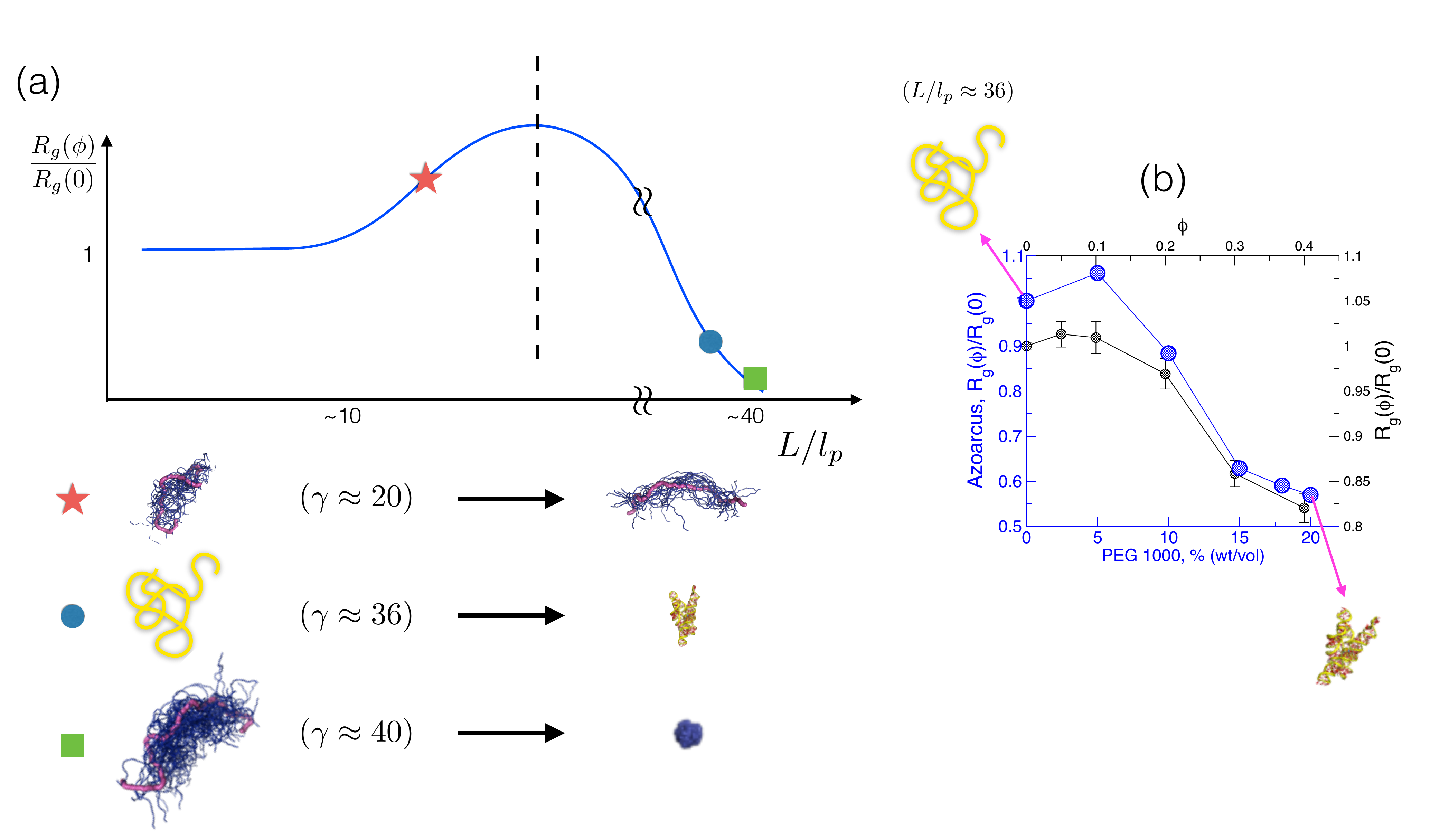}
\caption{\label{diagram}
(a) A schematic of the expected changes in the size of biopolymers in the milieu of \emph{E. coli}-like polydisperse crowding environment with $\phi=0.3$ as $L/l_p$ is varied.  
Depending on the parameter value $\gamma=L/l_p$, which characterizes the chain length and stiffness, the polymer undergoes swelling ($\gamma \geq 10$) or coil-to-globule transition ($\gamma\gg 10$).  
(b) PEG-induced compaction of \emph{Azoarcus} ribozyme in 0.56 mM-Mg$^{2+}$ ion solution (blue circles). For comparison we show in simulation (black circles) results for $R_g(\phi)$ changes in monodisperse SS crowding from \ref{Fig1-2}a. 
}
\end{figure*}

We provide arguments that when $\gamma$ exceeds a certain value the DNA-like polymer must undergo a transition from the swollen state to a globule. In other words, there must be a sharp coil-globule transition induced by the crowders. When $\gamma \gg 1$ chain stiffness is not that relevant and the polymer behaves like a flexible polymer. In this limit, using our results in ref.\cite{Kang15PRL} we predict that 
$\gamma$ has to exceed 40 to observe a genuine coil-globule transition. The simulation results and the physical arguments allows us to predict a rich dependence of $R_g$ in the {\it E. Coli.} environment (\ref{diagram}a).
\\

{\bf Insights into crowding effects on RNA. }
Recent experiments have examined the effects of polytehylene glycol (PEG) on the folding of a ribozyme \cite{Kilburn10JACS}. 
It has been argued that the impact of PEG can be understood based on the excluded volume effect. 
The SAXS experiments on \emph{Azoarcus} ribozyme with 195 nucleotides shows that $R_g$, near the midpoint of the Mg$^{2+}$ ion needed for the folding transition, initially increases before becoming compact (\ref{diagram}b). 
Folding of this RNA is also accompanied by decrease in the persistence length, which can be modulated by crowders. 
For \emph{Azoarcus} ribozyme $\gamma\approx 36$ where we have used $L\approx 195\times 0.55$ nm and $l_p\approx 3$ nm \cite{Caliskan05PRL}. 
If the theoretical prediction for the monodisperse SS in Figure 2a is correct then we expect a modest increase in $R_g$ as PEG (assumed to be sphereical) concentrations increase. 
The experimental data is in qualitative agreement with this expectation. 
It would be most interesting to examine the effects of polydisperse crowding agents on the complex problem of RNA folding to further some of our predictions.

\section{Conclusions}
In summary, using \emph{explicit} simulations of crowding particles, we predict multiple and unexpected scenarios for the effects of polydisperse crowding environment on the size and shape of a semiflexible polymer, which has served as a model for DNA and even RNA.  
Depending on the size, shape, and composition of the mixtures of crowding particles we find evidence for both compaction, and surprisingly dramatic increase in size as well. 
The results are of great relevance to the recent explosion of interest in the behavior of RNA \cite{Kilburn10JACS}, DNA \cite{Vasilevskaya95JCP,Chen2015PRL}, and proteins \cite{guzman2014JPCB,politou2015COSB} in macromolecular crowding conditions both \emph{in vitro} and \emph{in vivo}. 
The prediction that shape of chains, such as DNA, RNA, and F-actin \cite{frederick2008JMB}, can be dramatically altered in a polydisperse milieu can be tested in experiments. 

\section{Methods}
{\bf Model. }
To study the effects of crowding particles on a stiff chain, we used 
a coarse-grained model of WLC polymer ($N_m=300$), SS and SC crowding particles (Figure 1a). 
The length scales, $\sigma_{sph}$, $\sigma_{cyl}$, and $\sigma_m(\approx 3.18\text{ nm}$ for DNA) denote the size (diameter) of the SS, SC crowding particles and the monomer of WLC polymer, respectively. 
We set the aspect ratio of the SC to be 2 and $\sigma_{sph} =2^{2/3}\sigma_{cyl}=4\sigma_m$, so that the volumes of the individual SS and SC crowding particles are identical. 
In the WLC model, chain connectivity, with a fixed bond length, is maintained using a large spring constant connecting two consecutive beads. The bending rigidity of the chain was implemented by quadratic bond angle potential. We chose Weeks-Chandler-Andersen (WCA) potential for interactions between monomers, and $r^{-12}$ soft-sphere potential is employed for excluded volume interactions for crowder-crowder and crowder-polymer beads. 
\\

{\it Soft Spheres (SS)}
The energy function for the system consisting of the WLC and soft spherical crowders is,

\begin{equation}
E = E_{S} + E_{B} + E_{\mathrm{WCA}} + E_{R},
\label{eq:E}
\end{equation}

\begin{equation}
E_{S} = \sum_{i=1}^{N_{m}-1} K \frac{ \left( \left| \vec{r}_{i+1} - \vec{r}_{i} \right| - l_{0} \right)^{2} }{l_{0}^2},
\label{eq:ES}
\end{equation}

\begin{equation}
E_{B} = \sum_{i=1}^{N_{m}-1} G \left( \theta_{i} - \theta_{0} \right)^{2},
\label{eq:EB}
\end{equation}

\begin{equation}
E_{\mathrm{WCA}} = \sum_{i,j<N_{m}} \Theta \left( \frac{\sigma_{ij}}{r_{ij}} - 1 \right) \epsilon \left[ \left( \frac{\sigma_{ij}}{r_{ij}} \right)^{12} - \left(  \frac{\sigma_{ij}}{r_{ij}} \right)^{6}  \right],
\end{equation}

\begin{equation}
E_{R} = \sum_{i<j} \epsilon \left( \frac{\sigma_{ij}}{r_{ij}} \right)^{12} , \sigma_{ij} = \frac{\sigma_{i} + \sigma_{j}}{2}
\end{equation}
where $\sigma_{ij} = \frac{\sigma_{i} + \sigma_{j}}{2}$, $\Theta \left( x \right)$ is a Heaviside step function, $N_{m}$ is the number of monomers, $K$  ($>1000$ $k_{B}T/l_{0}^2$) is the spring constant with $l_0$ being the bond length. The bending rigidity constant is $G$, $\sigma_{i}$ is the diameter of a bead, $\theta_{i}$ is the angle between the monomer bond vectors $(\vec{r}_{i+1}-\vec{r}_{i})$ and $(\vec{r}_{i}-\vec{r}_{i-1})$, and $\epsilon$ is Lennard-Jones energy constant controlling the strength of the excluded volume interaction. The diameter of the SS, $\sigma_{sph} = 2 \sigma_{m}$ where $\sigma_{m}$ is the size of the monomer. We chose $N_{m}=300$ in our simulations.
\\

{\it Crowders as spherocylinders (SC). }
We model the spherocylindrical crowder by connecting five spherical crowders allowing for overlap 
(Figure 1a). Five beads in the anisotropic crowders are connected using $E_{S}$ and $E_{B}$ in Eqs. (\ref{eq:ES}) and (\ref{eq:EB}) with very large values for the spring and the bending rigidity constants ($K_{sp}$ and $G_{sp}$, the analogues of $K$ and $G$ in \ref{eq:ES} and \ref{eq:EB}) in order to maintain the cylindrical shape. We ignore excluded volume interaction between the beads within a particular cylindrical crowder, because the parameter $G_{sp}$ is sufficiently large. 
By choosing the diameter of the cylinder $\sigma_{cyl}=2.36\sigma_{m}$, the volumes of SS and SC crowders are identical ($\frac{1}{6}\pi \sigma_{sph}^3=\frac{1}{6}\pi\sigma_{cyl}^3 + \frac{\pi \sigma_{cyl}^2}{4}\times 2 \sigma_{cyl}$) (Figure 1a). 
The parameter values are given in Table S1. 
\\

{\it Mixture of SS and SC. }
To examine if the crowders of different shape have additive effects on the size of the WLC, we also  considered a system containing spheres and spherocylinders. We chose equimolar mixture containing $N_1 = N_2=N/2$ spheres and spherocylinders so that the total volume fraction of the crowders is $\phi=N_{1}\left(v_{SS}+v_{SC}\right) = N v_{SS}$ with  $v_{SS}=v_{SC}$. 
Thus, in a mixture $\phi_{SS}=\phi_{SC}=\phi/2$ where $\phi_{SS}$ and $\phi_{SC}$ are the fractions of volume occupied by SS and SC, respectively.
\\

{\it Modeling the E. coli environment. }
Using the approximate composition of the crowding particles in \emph{E. coli} in terms of sizes of crowders \cite{Roberts2011PLOSCOMPBIO} we mimic the cytoplasm as a mixture of spheres containing the three largest particles. 
They are the ribosome with radius $r_1=10.4$ nm, polymerases and other large complexes with average size $r_2=5.2$ nm, and smaller complexes where mean size $r_3\approx 2.6$ nm. 
The composition of the three classes of particles is 11 \%, 11 \%, and 8 \% respectively. 
These simulations provide a general framework for understanding the fate of stiff molecules in cell-like environment. 
\\

{\bf Simulation details. }
In order to obtain adequate sampling of the conformational space of the system, we used low friction Langevin dynamics (LFLD).   It can be shown rigorously, and has been confirmed in simulations, that the thermodynamic properties of the system do not depend on the choice of the friction coefficient \cite{Denesyuk11JACS}. 
In the LFLD, the diameter of the monomer $\sigma_{m}$, $\tau=(m \sigma_{m}^2/\epsilon)^{1/2}$, and $\epsilon$ were chosen as the units of length, time, and energy, respectively. The value of $\sigma_{m}$ suitable for DNA is $\approx3.18$ $\mathrm{nm}$. Friction coefficients, $\zeta_{m}$, for monomers and $\zeta_{c}$ for crowders were $\zeta_{m}=0.05$ $m \tau^{-1}$ and $\zeta_{c} = \zeta_{m} \sigma_{sph}/\sigma_{m}$  \cite{VeitshansFoldDes97} (Table S1). The duration of each trajectory ranges from $2\times10^{7}$ to $5\times10^{7}$ $\Delta t$ where $\Delta t=0.01$ $\tau$. 

Initially, a semi-flexible chain was placed in a simulation box without the crowders. We performed the LFLD for $10^{7}$ time steps to equilibrate the system. The crowding particles were added to generate a sample with $\phi=0.05$. Higher volume fractions were reached by inserting additional crowding particles to the simulation box.
Subsequently, Lennard-Jones interaction annealing (adiabatic increase of $\epsilon$) was carried out for $\phi\ge 0.3$ in order to improve the speed of equilibration, and to avoid catastrophic crashes during the insertion of particles. 
To be specific, at the beginning of annealing, we inserted the crowding particles at random positions by assigning $\epsilon=0.1$ $k_BT$ and $0.1 \Delta t$ for the integration time step of LFLD. 
First, the Lennard-Jones interaction parameter $\epsilon$ was increased by $0.05$ $k_{B}T$ for every $10^4$ time steps until it reaches $\epsilon=1.67$ $k_{B}T$. 
Next, we increased the simulation time step by $0.1 \Delta t$ for every $10^4$ time steps until the time step reaches $\Delta t$.

For purposes of efficient computation we devised a method (see also SI in Ref. \cite{Kang15PRL}) in which crowding particles are added on the fly and the volume of the simulation box is adjusted to keep the volume fraction constant.
During the simulations we adjusted the size of the simulation box according to the chain conformation at a given time to minimize the number of crowders. At every time step, we checked if the chain is enclosed in the simulation box. If any monomer and the boundary of the box is closer less than three times the average distance between the crowders, we resized the box and added crowding particles to the newly extended empty spaces. As a result of constantly resizing the box (a cuboid with changing dimensions) the number of crowders varies. The volume of the cuboid and the number of crowders are varied in such a way that $\phi$ is a constant. The average number of crowders in our simulations varies from 4000 to 8000 depending on $\phi$. 

The particular ensemble used in these simulation is not used frequently although it is discussed by Callen (see page 148 in \cite{CallenBook}). If the system consists of mono-disperse particles (generalization to multi-component system follows readily as explicitly shown in Ref\cite{CallenBook}) in which the number of particles ($N$) and volume ($V$) fluctuate, the independent variables conjugates to these two variables are chemical potential ($\mu$) and pressure ($p$). 
The thermodynamic potential in this ensemble is $\Omega(\mu,p,T)= U - TS + pV - \mu N = G-\mu N=0$ (follows from Euler relation), which means that variations in $N$ and $V$ do not change the thermodynamic potential $\Omega(\mu,p,T)$. 
This is precisely what is desired in these simulations. 

To ensure that the results do not depend on the choice of ensemble we also repeated the simulations in the canonical ensemble for the polydisperse case. As expected on theoretical grounds the results for $R_g$, $P(R_g)$, and the energy per particle (thermodynamic quantity) in the two ensembles are the same. The comparison is given in the Fig.S6.

In total, we generated 25 trajectories at each volume fraction to obtain statistical properties. We collected data for analysis after a minimum of  $10^{6}$ simulation time steps. 
\\

{\bf Acknowledgements.} This work was supported in part by the National Science Foundation (CHE 13-61946).

\bibliographystyle{jacs}
\bibliography{mybib1}
\clearpage 
 
\setcounter{figure}{0} 
\setcounter{equation}{0}
 
\makeatletter
\renewcommand{\theequation}{S\@arabic\c@equation}
\renewcommand{\thefigure}{S\@arabic\c@figure}
\renewcommand{\thetable}{S\@arabic\c@table}
\makeatother

\section{Supporting Information}

{\bf Multiple layered neighbor list. }
The large system size needed to reliably simulate the WLC chain in the presence of explicit crowders is computationally demanding. 
To To circumvent the system size problem, we develop a computer code that minimizes the number of operations to compute interaction potential. In order to acheive this goal, 
we devised and implemented the multiple layered neighbor list (MLNL) technique. 
To our knowledge, this methodology has not been used in simulations before. In conventional Verlet algorithm, the list of neighbors, which are the particles located within a cut-off distance, $R_c$, is created for each particle at the beginning of the simulation. When a pair-interaction is needed, we search only the neighbors instead of computing interactions between all pairs of particles. Thus, the computational cost decreases as $R_{c}$ gets smaller.
However, since the positions of the particles are constantly evolving, we have to update the neighbor list with a certain frequency. In conventional Verlet list, all neighbor lists are updated whenever the maximum displacement of any particle exceeds $R_{c}$. The frequency of updates increases as $R_{c}$ becomes smaller, thus increasing the computational costs for updating the neighbor list. The two competing demands (frequent update for small $R_{c}$ and infrequent update for computations involving larger number of interaction pairs) requires an optimal value of $R_{c}$.

MLNL is designed to reduce the computational costs for calculating interaction potentials and updating neighbor lists by using multiple numbers of neighbor lists. It consists of several neighbor lists each with a different cut-off distances, $R_{c}^{(1)}<R_{c}^{(2)}\cdots<R_{c}^{(n)}$. Interaction potentials are only calculated using the upper-most layer, which has the smallest $R_{c}=R_{c}^{(1)}$. This strategy minimizes the cost of calculating interaction potentials. 
When the maximum displacement of a particle exceeds $R_{c}^{(1)}$, instead of calculating the distance between all pairs of particles as in the conventional algorithms, we update the upper-most neighbor list using the neighbor list with $R_{c}^{(2)}$. Thus, the requirement of computing $O\left(N^{2}\right)$ interactions is avoided to a large extent by using this technique.
The disadvantage of the MLNL is that memory requirement can be quite large especially when the system size is large. For $N=300$, three layers suffice. With this choice we were able to perform converged simulations.
\\

\begin{figure}[t]
\begin{tabular}{c}
\includegraphics[width=0.8\columnwidth]{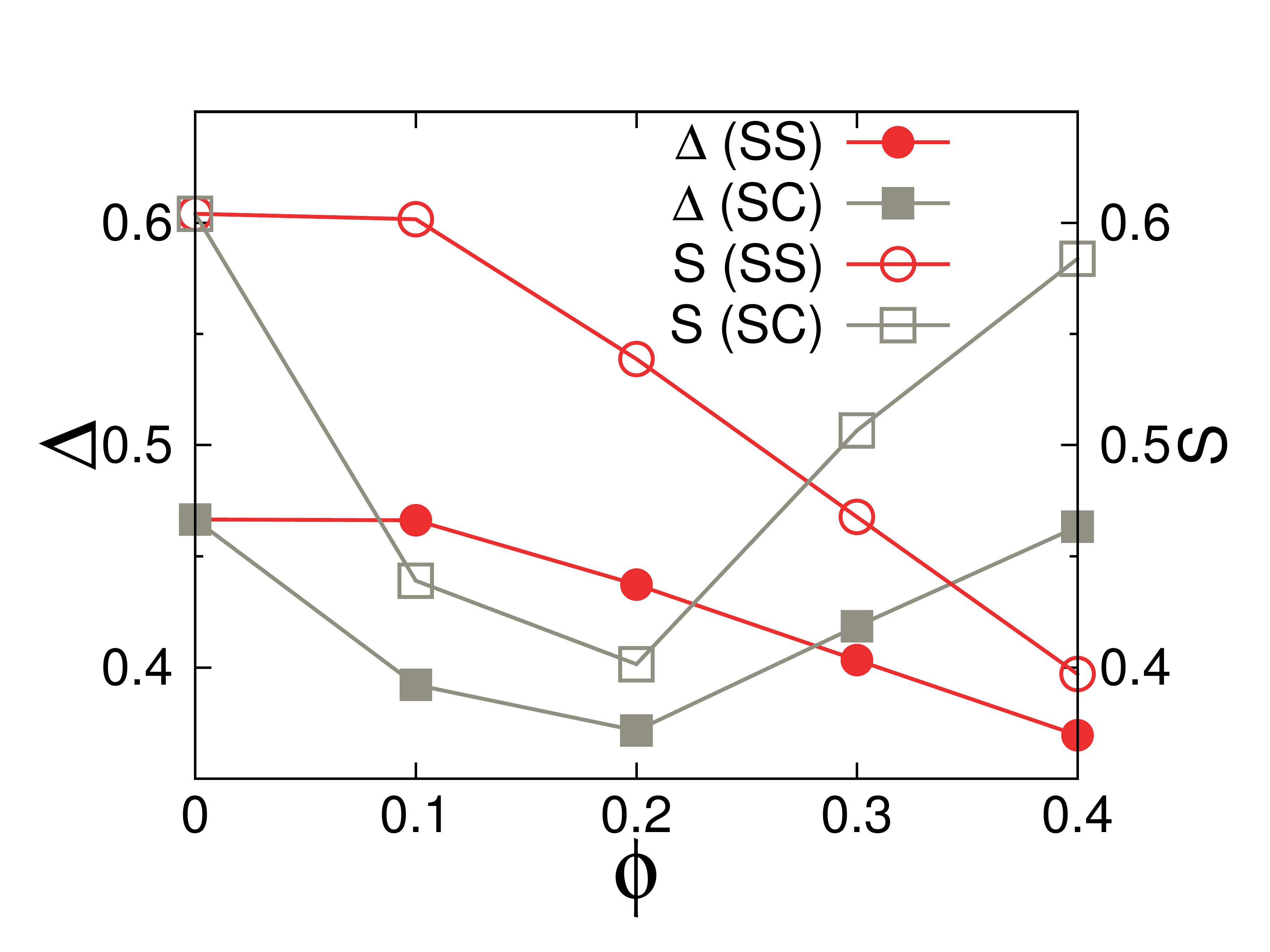}
\end{tabular}
\caption{
Asphericity ($\Delta$) and shape parameter ($S$) of WLC as a function of SS and SC volume fraction, $\phi$.  
\label{Shape}
}
\end{figure}

{\bf Shape anisotropy of polymer. }
Asphericity ($\Delta$) and shape parameters ($S$) are used to characterize the shape anisotropy of polymer. 
The asphericity parameter ($\Delta$) and a shape parameter ($S$). Both $\Delta$ and $S$, which are rotationally invariant, are defined using the inertia tensor,
\begin{equation}
T_{\alpha \beta} = \frac{1}{2 N^{2}} \sum_{i,j}^{N} \left( r_{i \alpha} - r_{j \alpha} \right) \left( r_{i \beta}-r_{j \beta} \right)
\end{equation}
where $r_{i\alpha}$ is the $\alpha\left(=x,y,z\right)$ component of bead $i$ of the WLC chain. The eigenvalues of $T_{\alpha \beta}$ are related to $R_{g}$ via $R_{g}^2 = \mathrm{Tr}(T)=\sum_{j} \lambda_{j}$. 
The anisophicity ($\Delta$) and shape ($S$) parameters are
\begin{equation}
\Delta = \frac{3}{2} \sum_{i=1}^{3} \frac{\left(\lambda_{i} - \bar{\lambda}\right)^{2}}{\left(\mathrm{Tr}(T)\right)^{2}}, \quad S=\frac{27 \prod_{i=1}^{3} \left( \lambda_{i}-\bar{\lambda}\right)}{\left(\mathrm{Tr}(T) \right)^{3}}
\end{equation}
where $\quad \bar{\lambda}=\frac{\mathrm{Tr}(T)}{3}$. 
For a globule, $\Delta=S=0$. Thus, we expect that if crowders induce compaction then $\Delta$ and $S$ should decrease monotonically as $\phi$ increases.

Compaction in $R_{g}$ for $\phi\leq0.3$ can also be correlated with shape changes. 
We calculated asphericity ($\Delta$) and shape parameter ($S$) as a function of crowder characteristics (Fig.\ref{Shape}). 
For spherical crowders, both $\Delta(\phi)$ and $S(\phi)$ decrease with increasing $\phi$, implying that the WLC becomes more spherical with increasing compaction. 
In contrast, there is a sharp increase in $\Delta(\phi)$ and $S(\phi)$ as the volume fraction of the SC crowders increases beyond $\phi\approx0.2$ (Fig.\ref{Shape}). 
The value of $\Delta(0.4) \approx \Delta(0)$ and $S( 0.4) \approx S(0)$, which shows that at both $\phi=0$ and high $\phi$ the WLC is an anisotropic ellipsoid ($S>0$). The elongation of the chain along the local nematic axis is also reflected in an increase in $R_g(\phi)$ when $\phi>0.2$.
\\

{\bf Analytical expression for the end-to-end distance distribution of WLC. }
The end-to-end distance distance distribution of WLC, obtained in Ref. \cite{HyeonJCP06},
\begin{equation}
P(r) = \frac{4\pi \mathcal{N}r^2}{\left(1-r^2\right)^{9/2}} \exp{\left[ -\frac{3}{4}\frac{t}{\left(1-r^2\right)}\right]}
\label{eq:TH}
\end{equation}
with $r=R_{ee}/L$, $t=L/l_{p}$, $\mathcal{N}=\frac{4 \pi^{-3/2} c^{3/2} e^{c}}{4+12c^{-1}+15 c^{-2}}$ and $c=\frac{3}{4} t$ where $L=(N_{m}-1)\sigma_m$ is the contour length was used to obtain numerical values of the persistence length, $l_p$.  
\\

{\bf Pair correlation function of polydispersed crowders. } 
Homogeneously distributed, the average separations ($D$) between crowders in cell lysate would be $D_X/\sigma_X=(4\pi/3)^{1/3}(1/2)\phi^{-1/3}=$1.68, 1.68, 1.87 for $X=1$, 2, and 3, respectively \cite{Kang15PRL}. In contrast to this expectation, the pair correlation between the polydisperse crowders of each size indicates local size ordering, a major population at $r/\sigma_X\approx 1$, which is more pronounced than in the monodisperse case (Fig.\ref{RDF_polydisperse}).
\\

\begin{figure}[t]
\begin{tabular}{c}
\includegraphics[width=1.0\columnwidth]{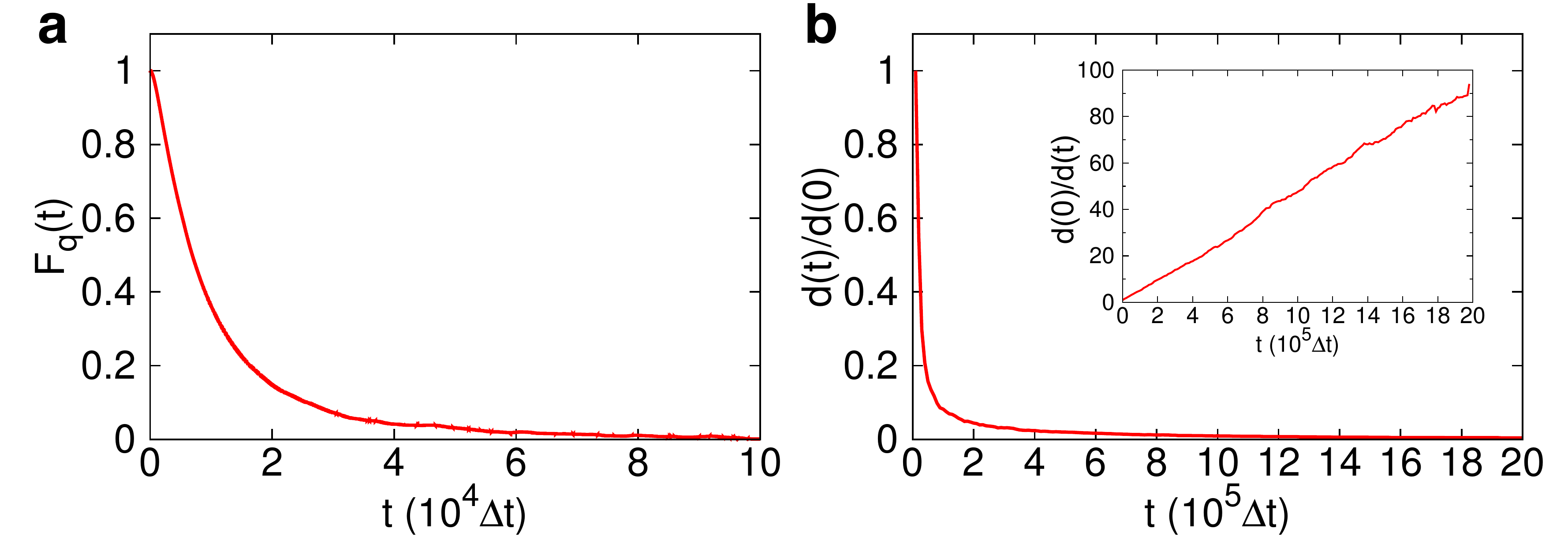}
\end{tabular}
\caption{\label{ergodicity}
{\bf a.} The time-dependence of the intermediate dynamic scattering function (Eq.\ref{eqn:Fq}). 
{\bf b.} Energy metric (Eq.\ref{eqn:metric}) as a function of $t$. We calculated these quantities to ensure that $F_q(t)\rightarrow 0$ and $d(t)/d(0) \rightarrow 0$ (or $d(0)/d(t)\sim t$ in the inset). Because the averages are computed using $t>10^7\Delta t$, whereas measure of ergodicity decay on $t\approx 10^4\Delta t$, the chain is fully equilibrated at $\phi=0.3$ in Figure 5.    
}
\end{figure}

{\bf Evidences of equilibration. }
Because the crowding particle sizes are large care must be taken to ensure that the system is well equilibrated.  
To provide explicit evidence that our computational results ($\phi=0.3$ case in Figure 3 in particular) are obtained from full equilibration, 
we evaluated multiparticle correlation function, which enables us to discern 
a glass-like dynamically arrested state from an fluid-like equilibrated state. 
The intermediate dynamic scattering function, which probes the density-density correlation in Fourier space 
at $q=|\vec{q}|=2\pi/r_s$ where $r_s$ is the position of the peak in the total pair distribution function, tells us how our system relaxes from an initial configuration: 
\begin{equation}
F_{\vec{q}}(t)=\frac{1}{N}\sum_{k=1}^Ne^{i\vec{q}\cdot(\vec{r}_k(t)-\vec{r}_k(0))}, 
\label{eqn:Fq}
\end{equation} 
where $\vec{r}_k(t)$ is the position of $k$-th monomer of our semiflexible chain at time $t$. 
As shown in Fig.\ref{ergodicity}a, $F_{\vec{q}}(t)\rightarrow 0$ on time scales that are nearly three orders of magnitude less than the length of trajectories used to affect equilibrium data. 
Thus, the memory of the initial configuration is fully erased in our simulations.  
In addition, as an alternative check for the equilibration, the energy metric for our system    \cite{ThirumalaiPRA90,Hyeon2012NatureChem,kang2013PRE},
\begin{equation}
d(t)=\frac{1}{N}\sum_{i=1}^N(\overline{E}_{\alpha,i}(t)-\overline{E}_{\beta,i}(t))^2
\label{eqn:metric}
\end{equation} 
where $\overline{E}_{\alpha,i}(t)=\frac{1}{t}\int_0^tE_{\alpha,i}(\tau)d\tau$ is the energy of particle $i$ averaged over time $t$ from the trajectory generated from two different initial condition $\alpha$ and $\beta$,   
shows that $d(t)/d(0)\rightarrow 0$ rapidly (Fig.\ref{ergodicity}b). 
Furthermore, $d(0)/d(t)\sim t$, a hallmark of ergodicity, is explicitly shown (Fig.\ref{ergodicity}b inset). 
The results show the stiff chain in the polydisperse crowding environment ergodically explore the conformational space, thus ensuring that the results in Figs.5 and 6 represent converged results.

\begin{widetext}
\begin{table*}[t]
\begin{ruledtabular}
\begin{tabular}{cccccccccccc}
$\sigma_{m}$&$\sigma_{c}$&$K$&$G$&$K_{sp}$&$G_{sp}$&$l_{0}$&$\theta_{0}$&$k_{B} T$&$\Delta t$&$\zeta_{m}$&$\zeta_{c}$\\
\hline
3.18nm&4$\sigma_{m}$&$1500\epsilon$&$4.91\epsilon$&$3000\epsilon$&$15\epsilon$&1.11 $\sigma_{m}$&$0$&0.6 $\epsilon$ & 0.01 $\tau$ & $0.05 m \tau^{-1}$ & $\zeta_{m} \left( \frac{\sigma_{c}}{\sigma_{m}}\right)$
\end{tabular}
\end{ruledtabular}
\caption{\label{table:parameters}Parameters characterizing the model. Lennard Jones energy constant $\epsilon$, the diameter of monomer $\sigma_{m}$ and $\tau=\sqrt{\frac{m\sigma_{m}^2}{\epsilon}}$ are used as the fundamental units for energy, length and time scales. $K$ and $G$ define the strength of bond and angle potentials (Eqs.~2 and 3);  $K_{sp}$ and $G_{sp}$ are the corresponding parameters for SC crowders. 
$l_{0}$ is a bond length between monomers of a chain, $k_{B}T$ is a temperature, $\Delta t$ is a simulation time step, $\zeta_{m}$ and $\zeta_c$ are the friction coefficients for monomer and crowders, respectively.}
\end{table*}
\end{widetext}

\begin{figure}[h]
\includegraphics[width=1.0\columnwidth]{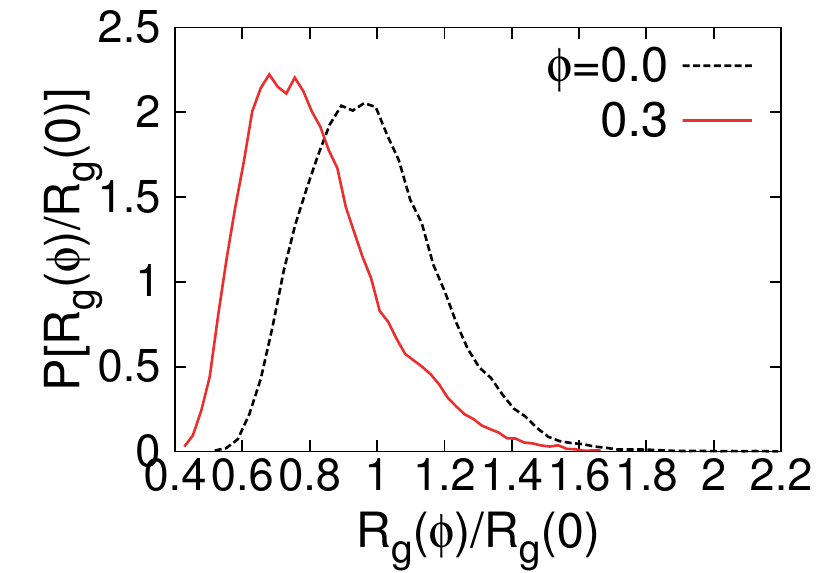}
\caption{\label{sawf_rgdist}
Distributions of gyration radius of SAW chain in the absence of crowders ($\phi=0.0$) and polydispersed crowder environment ($\phi=0.3$) mimicking the cytoplasmic condition of \emph{E. Coli}. Note that unlike WLC in the main text (Figure 5), SAW chain is compacted in the polydispersed crowding environment. }
\end{figure}

\begin{figure}[h]
\begin{tabular}{c}
\includegraphics[width=1.0\columnwidth]{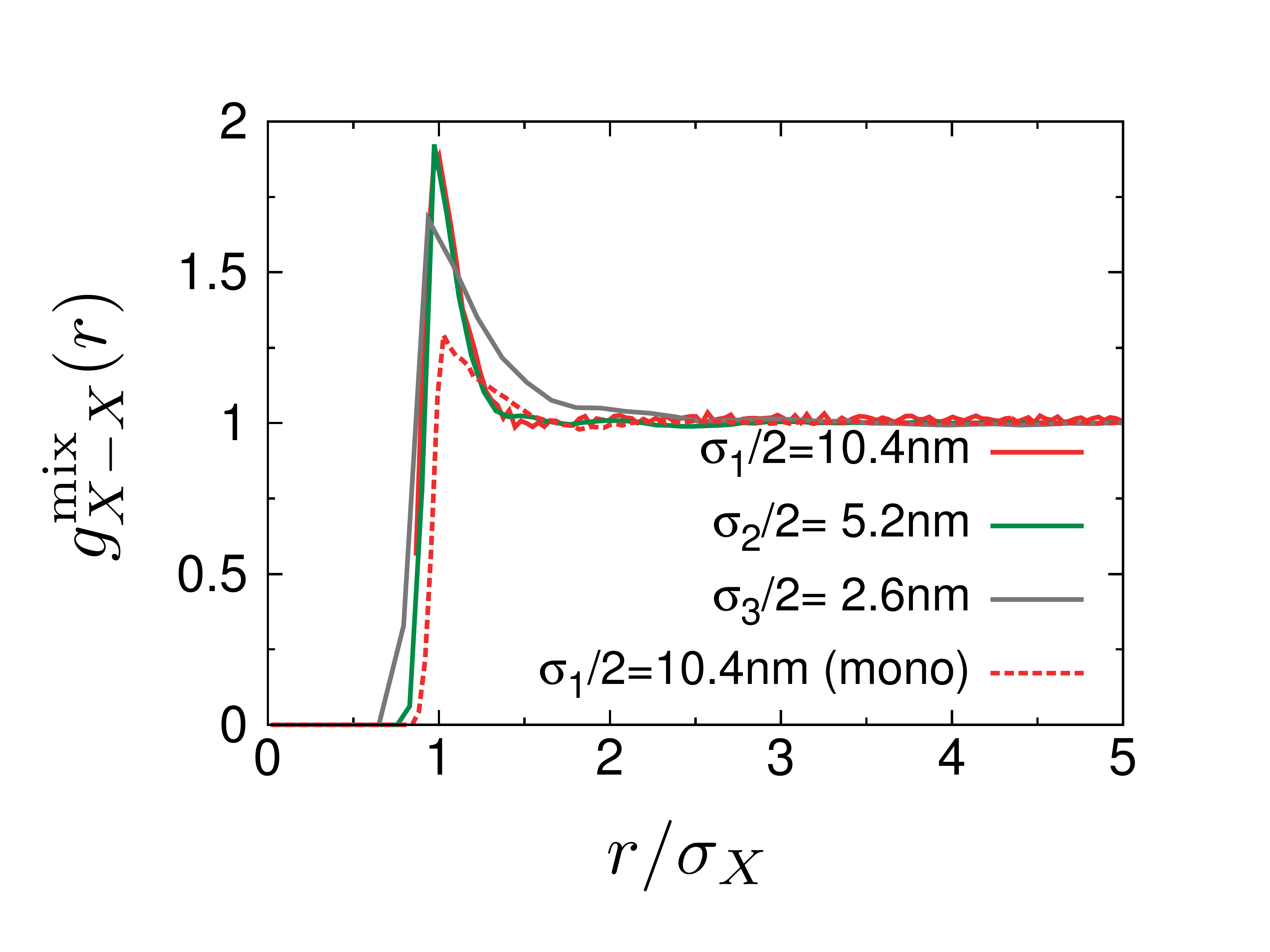}
\end{tabular}
\caption{\label{RDF_polydisperse}
Crowder-crowder pair correlation functions between crowders with the same kind in the polydisperse crowding solution.}
\end{figure}

\begin{figure}[t]
\begin{tabular}{c}
\includegraphics[width=1.0\columnwidth]{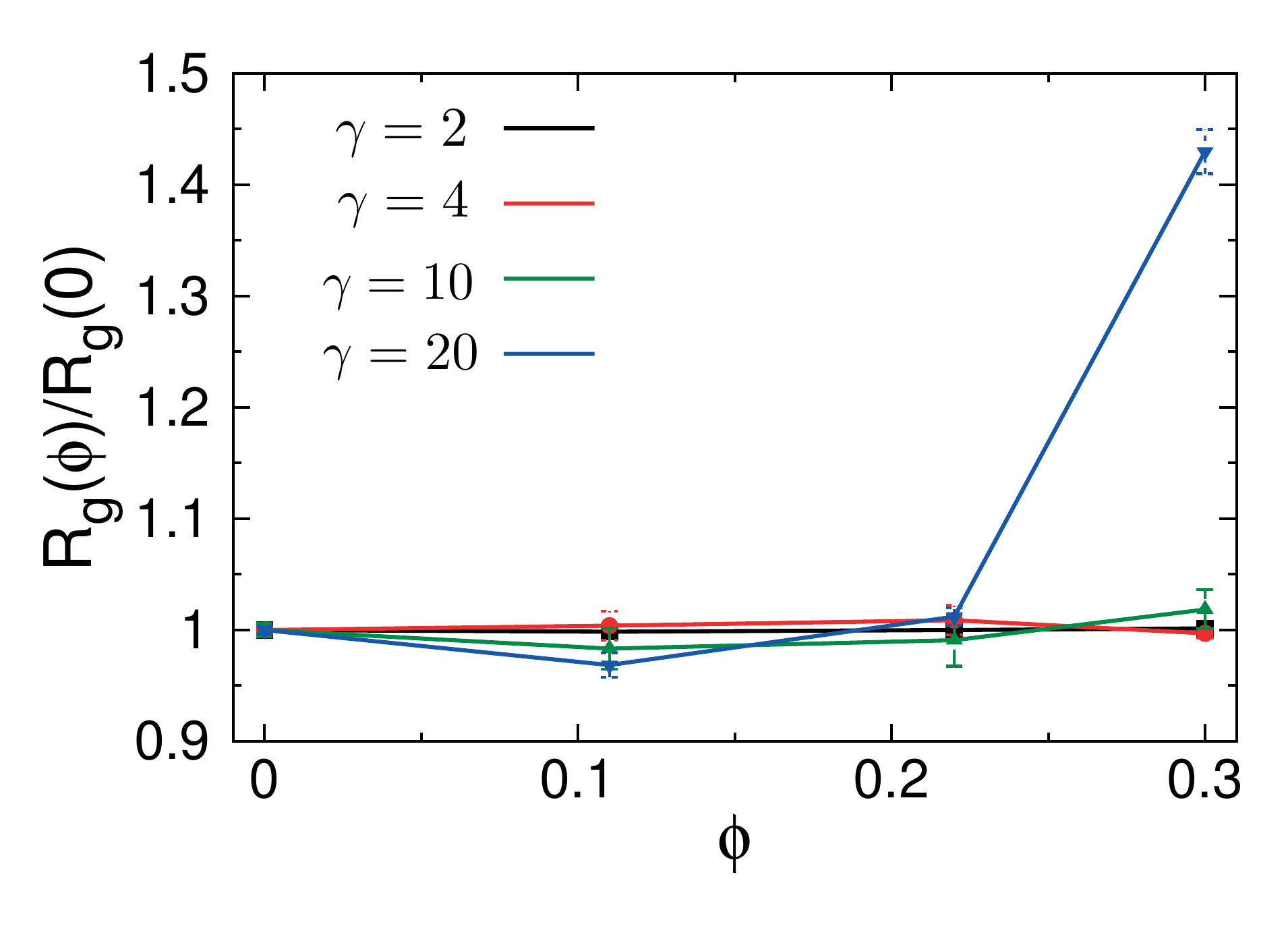}
\end{tabular}
\caption{\label{Loverlp_Ecoli}
Effect of polydisperse crowding environment on the size of polymers with different $\gamma=L/l_p$}
\end{figure}

\begin{figure}[t]
\begin{tabular}{c}
\includegraphics[width=1.00\columnwidth]{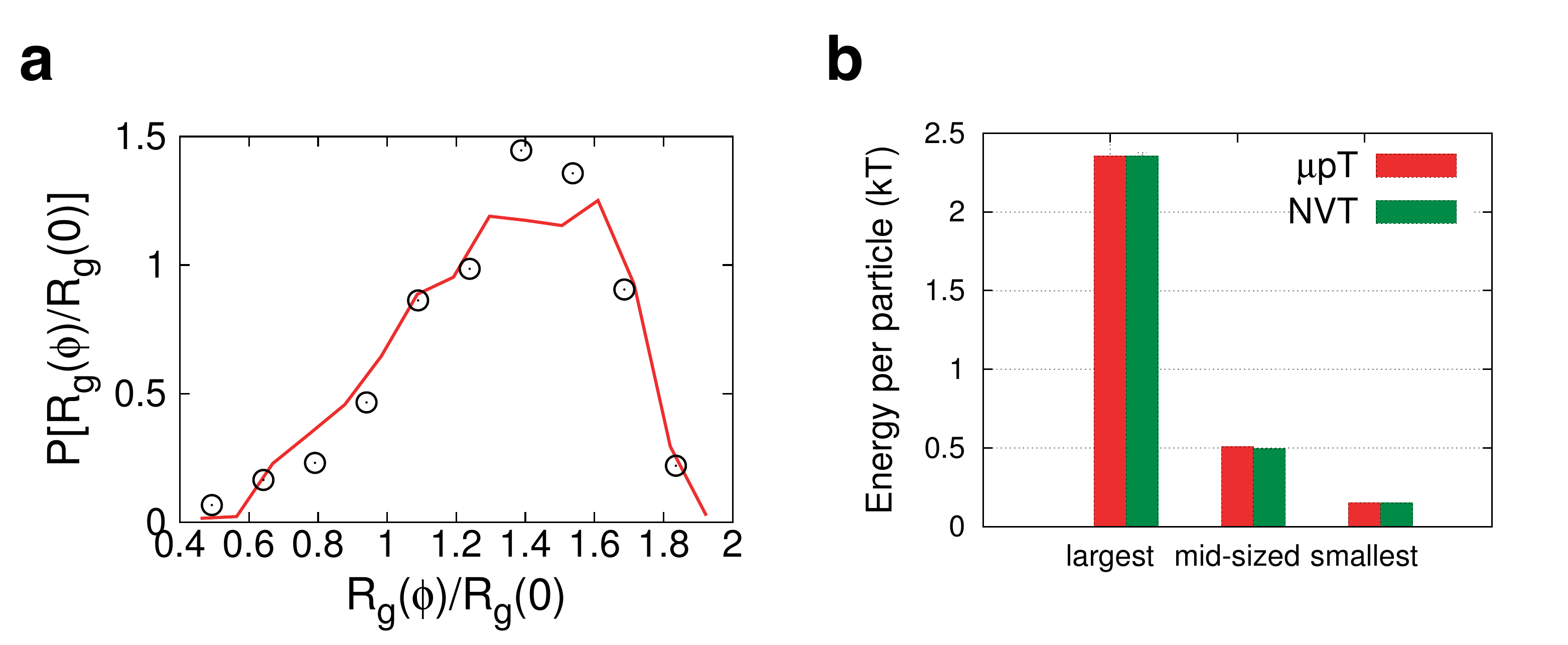}
\end{tabular}
\caption{\label{Loverlp_Ecoli}
Comparison of simulation results of WLC under polydisperse mixture environment in $\mu$PT and NVT ensembles. 
{\bf a.} $R_g$-distributions of WLC in NVT (circles) and $\mu$pT ensembles (red line). 
{\bf b.} Average energy per particle calculated for the two different ensemble. The value for each particle with different size is identical in the two ensembles, indicating that simulation results do not depend on the choice of ensemble. 
}
\end{figure}

\end{document}